# Integrated Differential Optical Shadow Sensor for Modular Gravitational Reference Sensor

**Andreas Zoellner[1,2], Eric Hultgren[1,2] and Ke-Xun Sun[2]**

[1]Department of Aeronautics and Astronautics, Stanford University, Stanford, CA 94305, USA

[2]Hansen Experimental Physics Laboratory, Stanford University, Stanford, CA 94305, USA

E-mail: zoellner@stanford.edu

**Abstract**. The Laser Interferometer Space Antenna (LISA) is a proposed space mission for the detection of gravitational waves. It consists of three drag-free satellites flying in a triangular constellation. A gravitational reference sensor is used in conjunction with a laser interferometer to measure the distance between test masses inside the three satellites. Other future space mission such as DECIGO and BBO also require a gravitational reference sensor. The Modular Gravitational Reference Sensor (MGRS) is being designed for these purposes and consists of two different optical sensors and a UV LED charge management system. The Differential Optical Shadow Sensor (DOSS) is one of the optical sensors and measures the position of a spherical test mass with respect to the surrounding satellite. This measurement is used for the drag-free feedback control loop. This paper describes the most recent, third generation of the experimental setup for the DOSS that uses a fiber coupled super luminescent LED, an integrated mounting structure and lock-in amplification. The achieved sensitivity is $10 \text{ nm/Hz}^{1/2}$ above 300 mHz, and $20 \text{ nm/Hz}^{1/2}$ for frequencies above 30 mHz.

## 1. Introduction to the Modular Gravitational Reference Sensor

Drag-free satellites are required for the Laser Interferometer Space Antenna (LISA) [1] as well as other fundamental physics missions [2] to achieve the requirements for orbit stability and position sensing set by the scientific mission goals. A gravitational reference sensor is the core instrument in a drag-free satellite where the satellite shields a free-floating test mass against external disturbances and follows the test mass through the use of a feedback control system.

The Modular Gravitational Reference Sensor (MGRS) [3-5] is a candidate sensor for these purposes and was presented to the LISA community in 2005. To lower the noise, the gap around the test mass in the MGRS is designed larger than 2 cm. This is beyond the sensitive range of capacitive sensors and thus the MGRS uses optical sensing.

A key concept for modularity is the use of two different optical sensors for the internal measurement: an interferometric sensor for the science measurement with a sensitivity on the order of picometers [6], and another sensor with a higher dynamic range and a sensitivity on the order of nanometers used mainly for the drag-free control loop. The latter is the Differential Optical Shadow Sensor (DOSS) addressed in this paper. The MGRS uses a single spherical test mass that can be spun up to shift disturbances associated with the test mass geometry out of the frequency band of interest for the scientific measurement. The DOSS will also be used on ground to characterize such spherical test masses by measuring the mass center offset and inertia properties.

## 2. Introduction to the Differential Optical Shadow Sensor

The DOSS, as part of the MGRS, is a sensor for measuring the position of a spherical test mass relative to a given reference system (the satellite in this case). The major difference between the shadow sensor and an interferometric measurement is that the measurement for the shadow sensor is based on light intensity rather than on phase information. This allows the use of non-coherent light sources for the shadow sensor. The dynamic range of the shadow sensor is large and only limited by the size of the detector and the beam waist.

Two light beams of equal intensity are tangent to and partially blocked by the test mass. The intensity of the two beams is measured and the difference between the two signals is taken. Thus, common mode noise that might exist in the signals is canceled out. The measurement principle is illustrated in Figure 1 for a one dimensional measurement with one pair of detectors. The drawing on the left shows the test mass, two detectors and the partially blocked light beams. The plot next to it shows a simulation of the signal measured by the left and right detector over the test mass displacement in the direction indicated by the arrow in Figure 1(a). The third line in the plot shows the difference between the left and right signal for which the sensitivity is doubled as compared to a single signal.

The DOSS combines several of these detector pairs in different layers and directions. As shown in the 3-D model in Figure 2, a tetrahedron configuration consisting of two detector pairs is used for a three dimensional measurement. The sensitive directions for the detector pair in the upper layer and the pair in the lower layer are perpendicular to each other and are, in case of the laboratory setup, used to measure a horizontal movement. Signals from the upper and lower layer are used to determine the movement in the third direction, the vertical one in this case.

The target sensitivity for the DOSS is 1 nm, which is the same as the requirement on the resolution of the inertial sensor for LISA [7].

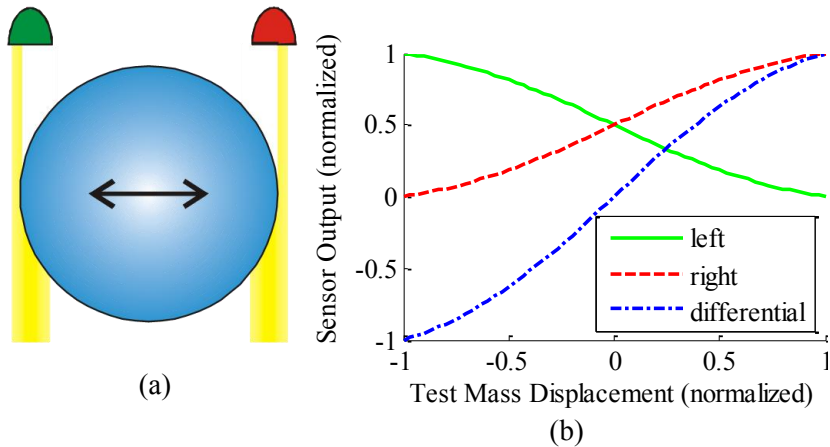
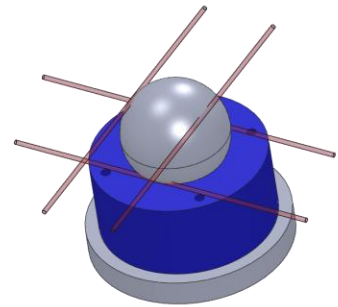

**Figure 1.** (a) Illustration of DOSS principle. (b) Sensor output for the left and right detector and differential signal over test mass displacement (normalized).

**Figure 2.** Illustration of tetrahedron configuration for three dimensional measurement.

## 3. History of the DOSS

We have conducted two earlier experiments for the DOSS. The first served as a proof-of-concept of the measurement principle, the second was a first step towards a three dimensional measurement with a spinning sphere and improved the sensor performance for low frequencies. Both experiments used a NPRO laser with a wavelength of 1064 nm as a light source and a single element InGaAs detector.

In the first experiment a custom-made one-dimensional flexure stage driven by a piezo actuator was used as a translational stage. The achieved noise level for the differential output from 2 Hz to 10 Hz is 2 nm/Hz$^{1/2}$ [8]. Below 2 Hz the noise level was dominated by a 1/f behavior. The upper frequency limit was set by the maximum sampling frequency of the data acquisition hardware.

The goal for the second experiment was to improve the sensitivity for low frequencies and to conduct a first measurement with a spinning sphere. Thus in the second setup, the translational stage was replaced with a commercial 3-axis piezo block from Thorlabs. An air-bearing was mounted on top of the piezo block to support the test mass. Furthermore lock-in amplification was used to overcome some of the low frequency noise. The noise level was better than 20 nm/Hz$^{1/2}$ for frequencies down to 60 mHz, the sensor performance was improved by one order of magnitude for frequencies below 100 mHz, i.e. the bandwidth of interest for LISA.

The first and second experiment mainly used detectors and optical components mounted on stainless steel posts for optics at a height of approximately 15 cm above the surface of the optical bench. Thus, torsional modes and bending modes of the posts caused misalignment of the sensors and introduced noise into the system. Furthermore, the experiment setup used mostly free space optics and thus noise due to air motion and temperature changes limited the performance of the shadow sensor.

The third generation of the DOSS experiment is a complete redesign that uses fiber optics and an integrated structure to mount the collimators and detectors. With its four detectors it will be able to perform a three dimensional differential measurement. A more detailed description of this setup follows in the next section.

## 4. Current generation DOSS

Several changes have been made compared to the earlier experiments in order to improve the performance at low frequencies and to reduce the complexity of the experiment with regard to its later use on a spacecraft. The third generation of the DOSS uses a fiber coupled super luminescent light emitting diode (SLED) as light source, quadrant photodiodes, a single rigid structure where all detectors and collimators are mounted to, as well as new amplifier electronics. In order to achieve lock-in amplification for all signal channels (eight per detector pair), the lock-in amplification introduced in the second generation was further developed and is now realized in software. As in the second experiment, a 3-axis piezo block and an air-bearing are used to hold the test mass. This allows for a three dimensional movement of the test mass as well as experiments with a spinning sphere. The schematic of the new setup is illustrated in Figure 3(a) where only one detector pair is shown to simplify the schematic and a recent photograph of the mounting structure around the test mass in the air-bearing is shown in Figure 3(b). Each discipline of optics, electronics, data acquisition and mechanics is explained in more detail in the following paragraphs.

### 4.1. Optics

In this experiment, a fiber coupled SLED from QPhotonics with a peak wavelength of about 1550 nm is used as a light source and split into several beams. The SLED is driven by a LED Controller from Thorlabs which has a modulation input that is connected to a function generator for the modulation of the light signal. The direct modulation of the beam power and the use of fiber optic components throughout reduces the number of optical components as compared to the earlier experiments that used a free space laser and the beam path exposed to air is shortened to a minimum in order to reduce the noise associated with air currents and temperature fluctuations. Presently, the controller limits the modulation frequency to about 150 kHz.

### 4.2. Electronics

A compact circuit board was designed to integrate a quadrant photodiode and four amplifiers that can be directly attached to the mounting structure. This reduces the electronic noise apparent in the earlier experiments due to long leads between the photodiode and the first amplifier stage. The amplifier is designed for a bandwidth that is a few times higher than the maximum modulation frequency.

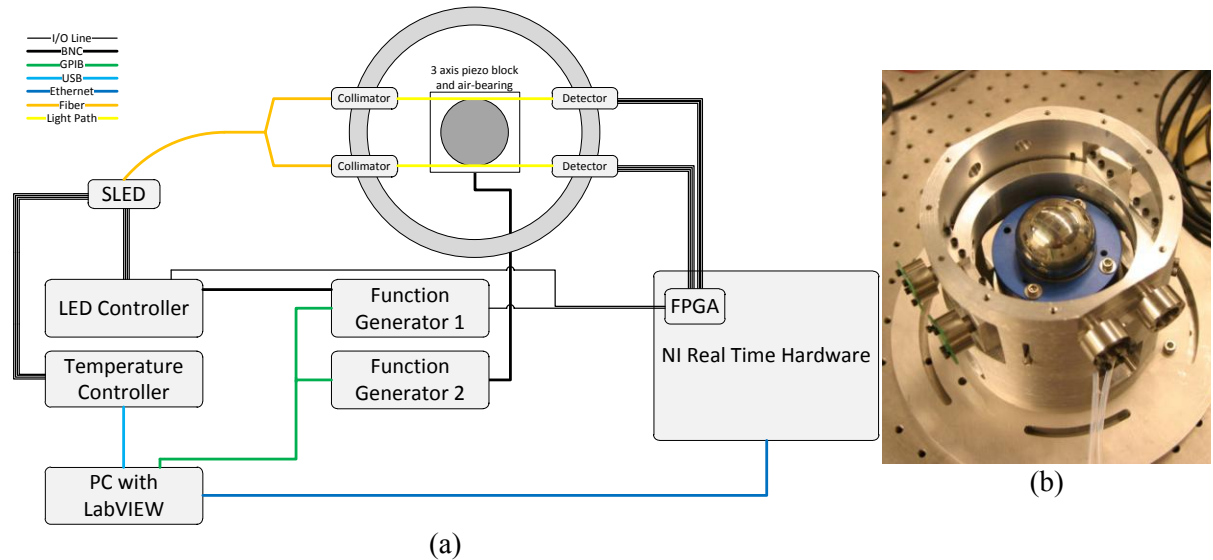

**Figure 3**. DOSS setup schematic (a) and photograph (b).

### 4.3. Data Acquisition and Signal Processing

The data acquisition is realized with a NI-PXI 7852R from National Instruments (NI) which integrates an FPGA with analog and digital inputs and outputs. It has 8 differential analog inputs with 16-bit resolution over a range of ±10 V and 96 configurable digital lines. The maximum analog sample rate is 750 kHz. The digital lines are used to enable and disable the output of the diode driver and to acquire the transistor-transistor logic (TTL) reference for the modulation signal. The analog inputs are used for the amplified photodiode signals.

The FPGA card is connected to an NI PXI 8108 which is an embedded controller running a real-time operating system from NI. The controller is used for the signal processing, mainly the lock-in amplification and filtering. As mentioned earlier, the lock-in amplification is realized in software. This makes it scalable and more channels can easily be added in the future without the need for new hardware.

As seen in Figure 3(a), all instruments are connected to a computer running LabVIEW and are integrated into a single Virtual Instrument (VI) for remote control of the experiment.

### 4.4. Mechanical Structure

The current experiment for the DOSS uses a cylindrical mounting structure of Aluminum 6061-T6 that is 152.4 mm (6 inches) in diameter with a wall thickness of 12.7 mm (0.5 inches). The upper plane of shadow sensors is integrated into this structure. A second plane of sensors is integrated into an annular ring 127 mm (5 inches) in diameter. With the integrated housing, the first bending mode, torsional motion, and radial expansion of the cylindrical cross-section ("breathing") cannot cause sensor misalignment.

## 5. Results

The data presented in this paper were taken with the setup described in section 4 but with one photodiode since the achieved mechanical tolerance for the mounting structure caused problems with the beam alignment for a differential measurement. For the data sets shown here, the modulation frequency was 2.3 kHz, the sampling frequency 50 kHz and the measurement time about 12 minutes. The lock-in algorithm was set up with a time constant of 250 ms and processed 10,000 data points at a time.

For calibration of the DOSS, the piezo block is moved horizontally with a sine wave input. The amplitude of the calibration signal is measured with a Michelson interferometer that uses a reference mirror at the piezo stage. For both data sets shown here the amplitude is 370 nm.

In the amplitude spectral density plot shown in Figure 4, two sets of data with different calibration signal frequencies are plotted. The calibration signal frequency of the first experiment is 100 mHz, the one of the second is 500 mHz.

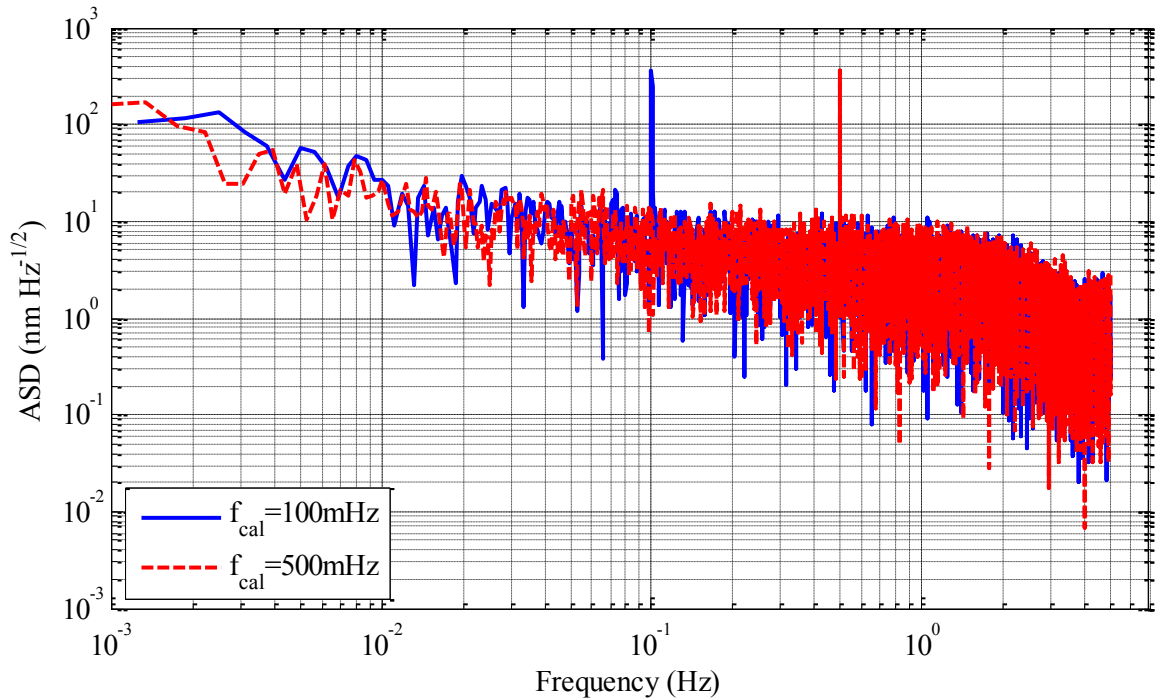

**Figure 4.** Amplitude spectral density of two position measurements with a calibration signal at 100 mHz and 500 mHz, respectively

In both experiments the noise floor that determines the sensitivity of the sensor is 20 nm/Hz$^{1/2}$ above 30 mHz and 10 nm/Hz$^{1/2}$ above 300 mHz. The spectral flatness is improved compared to previous experiments down to 10 mHz due to the use of lock-in amplification. Below this frequency the noise shows a 1/f behavior.

## 6. Conclusion

We have made substantial progress in integrating DOSS sensors with the MGRS structure to reduce mechanical instability, and in utilizing incoherent SLED light source to reduce scattering noise. Furthermore, we have used intensity modulation to code the DOSS beam and lock-in amplifiers to demodulate the signal. The net result is lower electronics noise in the low frequency band below 1 Hz, and improved detection spectrum flatness notably, till 30 mHz.

## 7. Future work

The ongoing work on the DOSS focuses on finalizing the setup for a three dimensional measurement with the new circuit boards mentioned above. Slight modifications to the data analysis software need to be made to account for the increased number of channels. That setup will then be used to perform experiments with a spinning sphere to measure the mass center offset and other parameters for a test mass model as well as observe phenomena such as the polhode motion [9].

Engineering wise, we will improve mechanical tolerances for DOSS source and sensor alignment and rigidity. At present the use of a manual mill is a limiting factor for the beam alignment in this fixed-beam setup. For future experiments, the beam alignment will be improved, e.g. by using a CNC mill, or adjustable beam alignment.

We will further place the entire MGRS and DOSS sensor structure in a vacuum enclosure. The air current below 1-2 Hz is recognized as main noise source.

The circuit boards used for the measurements described above have some limitations. Due to the limited space on the board, a quad-package amplifier had to be chosen and changes to the board could not be made easily after the photodiode was soldered in (and covered most of the components). Therefore new circuit boards are currently being designed. These will have special purpose amplifiers designed for low light photodiode applications as well as a second amplifier stage to convert the signal into a differential signal of the same range as the ADC inputs.

## 8. Acknowledgments

The authors gratefully acknowledge the support from NASA ROSES program, KACST and NASA Ames Research Center.References
[1] Bender P et al 1998 *LISA Laser Interferometer Space Antenna for the Detection and Observation of Gravitational Waves: Pre-Phase A Report* MPQ 233 2nd edn (Garching: Max Planck Institut für Quantenoptik)
[2] DeBra D B 2003 *Advances in Space Research* **32** 1221-26
[3] Sun K, Allen G, Buchman S, DeBra D and Byer R 2005 *Classical and Quantum Gravity* **22** S287-S296
[4] Sun K, Allen G, Williams S, Buchman S, DeBra D and Byer R 2006 *Journal of Physics: Conference Series* **32** 137-146
[5] Sun K et al 2006 *AIP Conference Proceedings* **873** 515-521
[6] Allen G 2009 *Optical Sensor Design for Advanced Drag-free Satellites* PhD thesis (Stanford University Stanford, CA, USA)
[7] The LISA International Science Team 2000 *LISA System and Technology Study Report* version 1.05
[8] Sun K et al 2009 *Journal of Physics: Conference Series* **154** 012026
[9] Conklin J W, Allen G, Sun K and DeBra D B 2008 *Journal of Guidance, Control, and Dynamics* **31** 1700-07